\documentclass[aps,pra,groupedaddress,twocolumn]{revtex4-2}

\usepackage{amsmath}
\usepackage{mathrsfs}
\usepackage{graphicx}
\usepackage[normalem]{ulem}
\usepackage{color}
\usepackage{gensymb}

\begin{document}

\renewcommand{\thefootnote}{\fnsymbol{footnote}}

\title{Static and dynamical properties of quadrupolar quantum droplets in quasi-2D condensates}

\author{Wei-qi Xia}
\thanks{These authors contributed equally to this work.}
\author{Xiao-ting Zheng}
\thanks{These authors contributed equally to this work.}
\author{Xiao-wei Chen}
\author{Gui-hua Chen}\email{Corresponding author. Email: cghphys@gmail.com}

\affiliation{School of Electronic Engineering $\&$ Intelligentization, Dongguan University of Technology, Dongguan, Guangdong 523808, China}

\date{\today}

\begin{abstract}
Quantum droplets, stabilized by beyond-mean-field effects, represent a novel and intriguing state of matter in quantum many-body systems. While previous studies have focused primarily on dipolar and contact-interacting systems, quadrupolar condensates remain relatively unexplored. In this work, we explore the formation, structural properties, and dynamical behaviors of quantum droplets in a two-component quadrupolar Bose–Einstein condensate confined to a quasi-two-dimensional geometry. By incorporating quadrupole–quadrupole interactions (QQIs) and Lee–Huang–Yang (LHY) corrections into an extended Gross–Pitaevskii equation, we systematically study both ground-state and vortex-state droplets. Analytical results obtained via the Thomas--Fermi approximation predict flat-topped density profiles and linear scaling between effective area and particle number. These predictions are corroborated by numerical simulations, which also reveal the saturation of peak density and chemical potential at large norm. Furthermore, vortex quantum droplets exhibit anisotropic elliptical morphologies due to the directional nature of QQIs, with their aspect ratios significantly tunable by varying the particle number and quadrupolar interaction strength. Collision dynamics demonstrate rich behavior modulated by velocity and topology: ground-state droplets transition from inelastic merging to quasi-elastic scattering and quantum penetration, while vortex droplets exhibit phase-induced repulsion, fragmentation, and topologically protected tunneling. These findings offer a comprehensive understanding of how higher-order interactions and quantum fluctuations govern the formation and stability of quadrupolar droplets. This work lays a theoretical foundation for experimental realization and opens new directions for exploring anisotropic quantum fluids, topological excitations, and applications in quantum sensing and simulation.
\end{abstract}


\maketitle

\section{Introduction}

Quantum droplets (QDs) arise as a distinct class of self-bound quantum states, stabilized by beyond-mean-field effects in ultracold many-body systems. Unlike classical liquids, these ultradilute quantum states are sustained by a subtle equilibrium between mean-field attraction and repulsive quantum fluctuations encapsulated in the Lee--Huang--Yang (LHY) correction \cite{Petrov2015, Lee1957}. Originally predicted by Petrov et al. for binary Bose mixtures near collapse thresholds \cite{Petrov2015}, the existence of QDs has been experimentally confirmed in potassium mixtures \cite{Cabrera2018, Semeghini2018} and heteronuclear condensates---first in a potassium--rubidium (K--Rb) mixture \cite{DErrico2019}, and later in a sodium--rubidium (Na--Rb) mixture \cite{Guo2021}---verifying their universal and robust character. Subsequently, dipolar quantum droplets, stabilized by the interplay of long-range dipole-dipole interactions (DDIs) and quantum fluctuations, have been observed in ultracold gases of dysprosium and erbium atoms, displaying pronounced anisotropy and self-bound behaviors distinct from their binary counterparts \cite{FerrierBarbut2016, Schmitt2016, Chomaz2016}.

Research on QDs has thus far predominantly concentrated on two classes of systems: binary condensates with short-range interactions and dipolar condensates governed by anisotropic, long-range interactions. In binary mixtures, the droplet formation mechanism relies on a delicate near-cancellation between interspecies attraction and intraspecies repulsion, critically stabilized by the LHY quantum correction \cite{Cabrera2018, Li2018, Kartashov2018, Dong2022, Yang2023}. Conversely, in dipolar condensates, QDs emerge from a near-balance between the repulsive contact interaction and the attractive component of the anisotropic DDI; the remaining net attraction is counteracted by the LHY correction, resulting in the formation of stable, strongly elongated droplets aligned along the polarization axis \cite{Baillie2017, FerrierBarbut2019, Cidrim2018, Xi2016, Bisset2021, Bland2022}. While these two paradigms have significantly advanced our understanding of quantum liquid phenomena, the stabilization and dynamics of more complex anisotropic interactions, particularly higher-order multipolar forces, remain largely unexplored. Recent experimental advances have revealed rich physics in dipolar supersolids and droplet crystals~\cite{Guo2019,Bottcher2019,Baillie2018}, highlighting the necessity of further exploring anisotropic interactions beyond the dipolar regime. For a comprehensive overview of recent advances and fundamental concepts of quantum droplets, including their formation mechanisms, stabilization effects, and distinctive quantum fluid properties, see the review by Luo et al.~\cite{Luo2021}.

In this context, quadrupole-quadrupole interactions (QQIs) represent a promising yet underinvestigated frontier. Quadrupoles, unlike dipoles, exhibit interactions characterized by intricate angular dependencies and shorter-range ($\sim1/r^5$) anisotropic potentials, offering rich possibilities for tuning interaction strengths and patterns \cite{Yi2000,Baranov2012}. Ultracold atomic and molecular gases possessing significant electric or magnetic quadrupole moments provide ideal platforms to explore novel quantum phases and self-localized states. Prior theoretical efforts have revealed that quadrupolar interactions can support anisotropic solitonic states elongated along specific axes determined by external fields \cite{Li2013, Huang2015, Chen2017}. Nevertheless, most earlier analyses of quadrupolar systems neglected crucial beyond-mean-field quantum corrections, specifically the LHY contribution, leaving open fundamental questions regarding the conditions necessary for stable quadrupolar quantum droplets and their associated dynamical properties. Recent theoretical and numerical studies in reduced-dimensional settings have highlighted that quantum corrections and anisotropic interactions can give rise to unique metastability phenomena and anisotropic droplet structures, motivating further exploration of these effects in quadrupolar quantum droplets~\cite{Li2017,Kartashov2019}. Specifically, recent work by Kartashov et al. has revealed that quantum droplet clusters exhibit intriguing metastable states, enriching our understanding of the underlying stability mechanisms and the role of quantum fluctuations in multipolar condensates~\cite{Kartashov2019}.

Furthermore, dimensionality substantially influences droplet formation and stability. Particularly in quasi-two-dimensional (quasi-2D) geometries, quantum fluctuations acquire distinctive density-dependent characteristics that profoundly alter stabilization mechanisms compared to three-dimensional settings. Recent works have demonstrated that combining anisotropic interactions, such as dipolar or quadrupolar forces, with dimensional reduction can yield qualitatively novel droplet phenomena, including enhanced anisotropy, modified stability thresholds, and novel collision dynamics absent in higher-dimensional systems \cite{Zheng2025, Yang2024}. 

In this context, two closely related studies deserve a brief comparison. Yang \textit{et al.} investigated two-dimensional quadrupolar droplets in binary mixtures with \emph{magnetic} quadrupole–quadrupole interactions (MQQI), reporting quasi-isotropic fundamental and vortex droplets (up to $S=4$) stabilized by the interplay of MQQI and LHY corrections \cite{Yang2024}. By contrast, Zheng \textit{et al.} analyzed \emph{electric} quadrupole–quadrupole interactions (EQQI), which are predominantly attractive on average, and demonstrated anisotropic droplets elongated along the polarization axis together with characteristic collision dynamics \cite{Zheng2025}. Compared with these works, the present study treats a symmetric two-component condensate with effective quadrupolar interactions reduced to a single-component eGPE framework. We emphasize the emergence of elliptical anisotropy, identify a finite norm threshold for stable vortex states, and map collisional outcomes (merging, right-angle deflection, tunneling, fragmentation). This focused contrast clarifies that earlier results highlighted isotropic versus strongly anisotropic behaviors depending on the underlying quadrupolar mechanism, while our analysis extends the scope by delineating quadrupolar droplets as a distinct self-bound phase governed by higher-order multipolar forces.

Motivated by these gaps in existing literature, we perform a systematic theoretical and numerical investigation of quantum droplets stabilized by QQIs in a two-component quadrupolar Bose-Einstein condensate (BEC) confined to a quasi-2D geometry. Our primary objectives are threefold: (i) to elucidate how QQIs, in concert with LHY quantum corrections, stabilize droplets in quasi-2D systems; (ii) to quantify the influence of quadrupolar interaction strength on droplet structural properties, including density distribution, chemical potentials, and effective areas; and (iii) to explore droplet dynamics, with a particular focus on vortex-state formation and collision behaviors under varying conditions. Specifically, we address crucial scientific questions: Under what parameter regimes can stable fundamental and vortex quantum droplets exist? How do QQIs quantitatively influence droplet morphology and stability compared to known binary and dipolar systems? Moreover, how do anisotropic interactions and dimensional effects manifest in collision dynamics and stability of vortex structures? Clarifying these points will significantly deepen our understanding of multipolar quantum systems and provide a robust theoretical foundation guiding future experimental realizations and practical applications of quadrupolar quantum droplets. 

The remainder of this paper is organized as follows. Section II outlines the theoretical model of the two-component quadrupolar BEC, which includes the long-range quadrupole–quadrupole interaction and beyond-mean-field LHY correction. In Section III, we perform a systematic theoretical analysis of the stationary states using the Thomas--Fermi (TF) approximation. Section IV presents numerical investigations of the stationary properties and stability of quantum droplets, covering both ground-state and vortex configurations. Section V is devoted to the collision dynamics of droplets under various initial conditions. Finally, Section VI summarizes the main results and discusses their implications for future theoretical and experimental developments in the field.

\section{Theoretical model for quadrupolar quantum droplets}

To address the outlined objectives, we construct a theoretical model that accurately captures the essential physics of two-component quadrupolar BECs in a quasi-2D setting. The system consists of molecular species possessing intrinsic electric quadrupole moments. A strong harmonic trapping potential tightly confines the condensate along the axial ($z$) direction, effectively restricting its dynamics to the transverse ($x$–$y$) plane. This strong confinement suppresses excitations in the axial direction and thus justifies the use of the quasi-2D approximation \cite{Bisset2016, Petrov2016}.

\begin{figure}[htbp]
	\centering
	\includegraphics[scale=0.45]{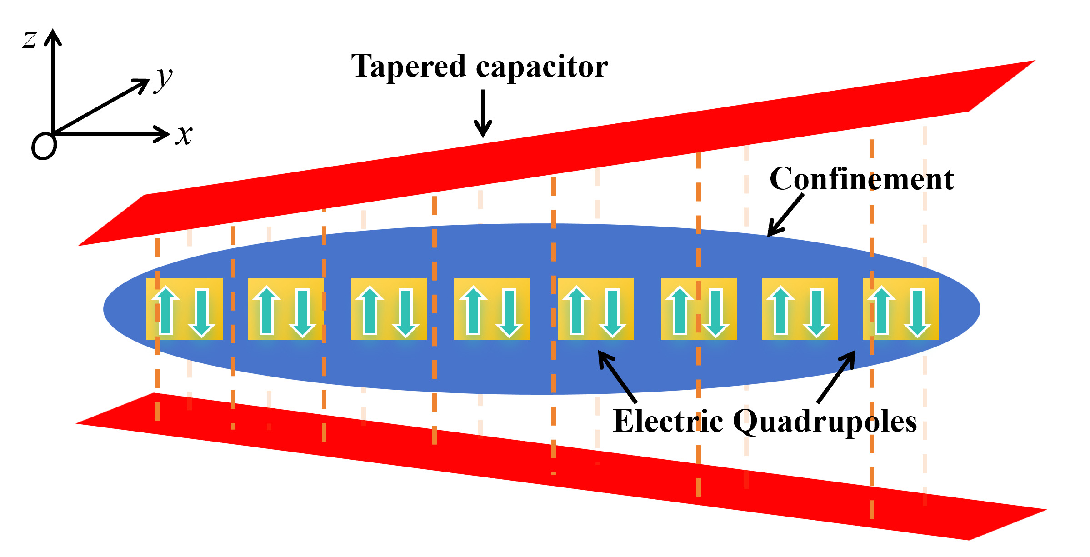}
	\caption{(Color online) Schematic illustration of the proposed experimental setup for realizing quadrupolar quantum droplets in a quasi-2D two-component Bose–Einstein condensate. Molecules with intrinsic electric quadrupole moments are uniformly aligned along the vertical $z$-axis by a spatially varying external electric field generated by a tapered capacitor. The system is tightly confined in the axial direction, resulting in effective two-dimensional dynamics in the $x-y$ plane. Each quadrupole is modeled as a dipole–antidipole pair (green up and down arrows), and the tunability of the QQI is achieved via the spatial modulation of the field gradient. This configuration allows for the exploration of self-bound quantum droplet states stabilized by the interplay between long-range attractive QQIs and repulsive quantum fluctuations. Similar mechanisms involving field-induced long-range interactions and spin-orbit coupling have been shown to support stable and even excited solitonic states in other two-dimensional condensate systems~\cite{Huang2018,Jiang2016}.} \label{Exp_Setup}
\end{figure}

Figure~\ref{Exp_Setup} illustrates the physical setup. The molecular quadrupoles, represented as dipole–antidipole pairs, are uniformly polarized along the $z$-axis by a spatially modulated electric field produced by a conical capacitor. This modulation enables controlled tuning of the quadrupole–quadrupole interactions (QQIs) within the $x$–$y$ plane, making it possible to explore anisotropic quantum effects. The proposed configuration is consistent with earlier theoretical schemes for manipulating quadrupolar interactions \cite{Huang2015, Zheng2025, Yang2024}.

The dynamics of a dilute, ultracold quantum gas at extremely low temperatures are accurately described by the Gross–Pitaevskii equation (GPE), extended here to include the critical beyond-mean-field quantum fluctuations (LHY corrections), essential for droplet stabilization. After appropriate nondimensionalization of space, time, and interaction parameters, the extended Gross–Pitaevskii equations (eGPEs) for the two-component quadrupolar condensate confined to a quasi-2D geometry take the following dimensionless form:

\begin{align}\label{Eq:GPE1}
	\left\{
	\begin{aligned}
	i\frac{\partial\psi_1}{\partial t} &= -\frac{1}{2}\nabla^2\psi_1 \\ & + K\psi_1\int R(\mathbf{r}-\mathbf{r}')\left(|\psi_1(\mathbf{r}')|^2+|\psi_2(\mathbf{r}')|^2\right)\mathrm{d}\mathbf{r}' \\ & + (g_{11}|\psi_1|^2 - g_{12}|\psi_2|^2)\psi_1 \\ & + \Gamma(|\psi_1|^2+|\psi_2|^2)\psi_1\ln(|\psi_1|^2+|\psi_2|^2), \\ 
	i\frac{\partial\psi_2}{\partial t} &= -\frac{1}{2}\nabla^2\psi_2 \\ & + K\psi_2\int R(\mathbf{r}-\mathbf{r}')\left(|\psi_1(\mathbf{r}')|^2+|\psi_2(\mathbf{r}')|^2\right)\mathrm{d}\mathbf{r}' \\ &+ (g_{22}|\psi_2|^2 - g_{21}|\psi_1|^2)\psi_2 \\ & + \Gamma(|\psi_1|^2+|\psi_2|^2)\psi_2\ln(|\psi_1|^2+|\psi_2|^2), 
	\end{aligned}
	\right.
\end{align}
where $\psi_{1,2}(\mathbf{r}, t)$ represent the condensate wavefunctions of the two components, and $\mathbf{r}=(x,y)$ denotes the planar coordinates. The kinetic energy term $-(1/2)\nabla^2$ accounts for quantum pressure associated with the spatial curvature of the wavefunction within the plane. The coupling constants $g_{11}$ and $g_{22}$ are positive and represent short-range repulsive intra-component interactions, while $g_{12}=g_{21}>0$, appearing with a negative sign in the coupled equations, describe attractive inter-component interactions. The parameter $K$ quantifies the strength of QQIs, adjustable by tuning the electric quadrupole moment of the particles.

The second term on the right-hand side incorporates the nonlocal, anisotropic QQIs characterized by the kernel function \cite{Huang2015}: 

\begin{equation} 
	R(\mathbf{r}-\mathbf{r}') = \frac{1-5\cos^2\theta}{\left[b^2+(\mathbf{r}-\mathbf{r}')^2\right]^{5/2}}, 
\end{equation} 
where $\theta$ denotes the angle between the vector connecting quadrupoles and the dipole-antidipole axis within the quadrupole (chosen here along the $x$-axis), and $b$ is a cutoff parameter representing the effective transverse confinement scale, ensuring convergence of the integral at short distances. The selection of $b=1$ establishes a physically meaningful reference length scale, regularizing the interaction potential and preserving essential anisotropic properties.

The last term explicitly includes the LHY quantum correction specifically adapted for quasi-2D systems, introducing a density-dependent logarithmic term that provides a repulsive quantum pressure critical for droplet stability against collapse induced by attractive QQIs \cite{Shamriz2020}.

To reduce the complexity of the coupled system while preserving the essential physics of droplet formation, we focus on the symmetric scenario characterized by equal masses and identical intra- and inter-component contact interactions, i.e., $g_{11}=g_{22}=g_{12}$, along with equal wavefunction amplitudes $\psi_1 = \psi_2 = \psi/\sqrt{2}$. This assumption is physically justified for mixtures of identical particles in different internal states (e.g., spin components), and allows for an effective reduction of the coupled extended Gross–Pitaevskii equations (eGPEs) to a single-component form. Under this assumption, the governing equation becomes:

\begin{equation}\label{Eq:GPE2} 
\begin{split}
	i\frac{\partial\psi}{\partial t} = & -\frac{1}{2}\nabla^2\psi + K\psi\int R(\mathbf{r}-\mathbf{r}')|\psi(\mathbf{r}')|^2 \mathrm{d}\mathbf{r}' 
	\\ & + \Gamma|\psi|^2\psi\ln(|\psi|^2). 
\end{split}
\end{equation}

To facilitate analysis and computations, we nondimensionalize the equations by measuring time and length in the units $t_0=1/\Gamma$ and $\ell_0=1/\sqrt{\Gamma}$, respectively, and by defining the dimensionless quadrupolar strength $\kappa \equiv K/\Gamma$. Writing $\tilde t = t/t_0 = \Gamma t$ and $\tilde{\mathbf r} = \mathbf r/\ell_0 = \sqrt{\Gamma}\,\mathbf r$, and then dropping tildes for brevity, the normalized eGPE takes the following form:
 
\begin{equation}\label{Eq:GPE3} 
\begin{split}
	i\frac{\partial\psi}{\partial t} = & -\frac{1}{2}\nabla^2\psi + \kappa\psi\int R(\mathbf{r}-\mathbf{r}')|\psi(\mathbf{r}')|^2 \mathrm{d}\mathbf{r}' \\ 
	& + |\psi|^2\psi\ln(|\psi|^2), 
\end{split}
\end{equation} 
with dimensionless QQI strength $\kappa$ measuring the relative magnitude of quadrupolar interactions versus quantum fluctuations. It is worth emphasizing that, in contrast to single-component dipolar or quadrupolar condensates—where strong long-range interactions are typically required to counteract the contact interaction and render the LHY correction significant—our two-component framework allows for an alternative strategy. By carefully balancing the intra- and inter-component contact interactions, one can effectively suppress the net mean-field nonlinearity. This reduction makes it possible to observe the direct competition between the quadrupolar interaction and the LHY term, even when the quadrupolar strength is moderate. As such, we choose the dimensionless quadrupolar strength $\kappa$ within the range $[0, 0.5]$, which enables a systematic exploration of droplet formation and evolution under experimentally accessible conditions.

The dimensionless Hamiltonian associated with the eGPE, describing the total energy, is expressed as:

\begin{equation}\label{Eq:energy}
\begin{split}
	E=&\frac{1}{2}\int \mathrm{d}\mathbf{r}\bigg[|\nabla\psi|^2+\kappa|\psi|^2\int R(\mathbf{r}-\mathbf{r}')|\psi(\mathbf{r}')|^2 \mathrm{d}\mathbf{r}' \\ 
	& +|\psi|^4\ln\frac{|\psi|^2}{\sqrt{e}}\bigg],
\end{split} 
\end{equation}
with the conserved norm (total particle number) given by:

\begin{equation} 
	N=\int |\psi(\mathbf{r})|^2 \mathrm{d}\mathbf{r}. 
\end{equation}

This comprehensive theoretical model rigorously encapsulates the essential physical interactions and quantum effects, providing a robust basis for exploring novel phenomena driven by anisotropic QQIs in two-component quadrupolar quantum droplets.

While the present model is formulated in dimensionless units and explores a regime where the dimensionless quadrupolar strength $\kappa$ ranges up to $0.5$, it is important to emphasize that the underlying physical ingredients are within reach of current experimental techniques. In particular, the realization of ultracold gases of polar molecules such as RbCs \cite{Molony2014}, NaK \cite{Will2016}, or NaCs \cite{Lam2022} enables access to strong long-range interactions, including electric quadrupole-quadrupole forces, by aligning molecular axes using external electric field gradients or spatially structured field configurations. 

Representative molecular scales help contextualize the quadrupolar regime discussed here. For instance, rovibrational-ground-state RbCs has a body-fixed permanent electric dipole moment $d_0 \simeq 1.23~\mathrm{D}$ measured by Stark spectroscopy, whereas typical electric quadrupole moments of simple diatomics are on the order of $|\Theta_{zz}|\sim 1~ea_0^2$ (one atomic unit); for example, $\mathrm{O}_2$ has $\Theta \approx -1.03\times10^{-40}\,\mathrm{C\,m^2} \approx -1.22~ea_0^2$ (cf.~Refs.~\cite{Somogyi2021,Couling2024}). In our quasi-2D nondimensionalization, the effective quadrupolar coupling enters through $\kappa \equiv K/\Gamma$; using $|\Theta_{zz}|$ in the above range together with a micrometer-scale transverse length yields $\kappa$ values consistent with those explored here ($\kappa \approx 0.02$–$0.10$).

We also note that, in the absence of dc electric fields, freely rotating polar molecules have a vanishing space-fixed dipole expectation so that long-range dipole–dipole forces average out, making quadrupolar interactions the leading anisotropic term; additionally, microwave-dressing protocols can suppress effective dipolar couplings while maintaining collisional stability (cf.~Refs.~\cite{Walraven2024,Anderegg2021,Schindewolf2022}). These considerations justify the quadrupole-dominant regime addressed by our model and numerical analysis in quasi-2D geometries.

The logarithmic quantum fluctuation term (LHY correction) incorporated in our model is derived under a quasi-2D confinement and has been successfully used in the theoretical description of dipolar and contact-interacting droplets \cite{Petrov2016, Bisset2016}. This term originates from seminal work on stabilizing Bose-Bose mixtures \cite{Petrov2015} and builds on comprehensive theoretical frameworks for trapped Bose gases \cite{Dalfovo1999}. Although the actual value of $\kappa$ in experimental systems may exceed the regime studied here, our choice allows for a controlled theoretical exploration of the competition between attractive quadrupolar interactions and repulsive quantum fluctuations. This setting serves as a minimal yet physically relevant framework for capturing the qualitative emergence and stability of self-bound droplet states with anisotropic interactions, and can be refined in future studies by mapping specific experimental parameters to the dimensionless formulation.

In the following sections, we employ the TF approximation to obtain analytic baselines for key droplet properties, and then validate and refine these predictions with full eGPE simulations. We separately consider ground-state and vortex-state droplets, and finally examine their stability and collision dynamics.

\section{Analytical framework: Thomas--Fermi approach}

Both ground-state and vortex-state QDs can be studied by solving the eGPE with the stationary-wavefunction ansatz

\begin{eqnarray}\label{Eq:WF}
	\psi(\mathbf{r}, t) = \phi(r) e^{-\mathrm{i}\mu t + \mathrm{i}S\theta},
\end{eqnarray}
where $\phi(r)$ is the radial profile of the stationary wavefunction, $\mu$ is the chemical potential, and \( S \) represents the topological charge (vorticity) of the droplet. Substituting Eq. (\ref{Eq:WF}) into Eq. (\ref{Eq:GPE3}), one obtains the radial equation governing the wavefunction:

\begin{equation}\label{Eq:Stationary}
\begin{split}
	\mu\phi=&-\frac{1}{2}\bigg(\frac{\mathrm{d}^{2}\phi}{\mathrm{d}r^{2}}+\frac{1}{r}\frac{\mathrm{d}\phi}{\mathrm{d}r}-\frac{S^{2}}{r^{2}}\phi\bigg) \\ 
		& +\kappa\phi(r)\int R(\mathbf{r}-\mathbf{r}^{\prime})|\phi(r^{\prime})|^{2}\mathrm{d}\mathbf{r}^{\prime}+\phi^{3}\ln\big(\phi^{2}\big).
\end{split}
\end{equation}

Under equilibrium conditions, QDs exhibit nearly flat-topped density profiles, thus justifying the use of the TF approximation, in which the kinetic energy term in Eq. (\ref{Eq:GPE3}) is neglected. Within the TF approximation, the density distribution becomes spatially uniform, i.e., $n(\mathbf{r}) = |\psi(\mathbf{r})|^2 = \mathrm{const}$. Employing this constant-density assumption in the energy functional given by Eq. (\ref{Eq:energy}), we obtain the total droplet energy as

\begin{eqnarray}\label{Eq:E}
\begin{split}
	E=&\pi S^2n\ln\left(\frac{r_\mathrm{out}}{r_\mathrm{in}}\right) \\
	& +\frac{1}{2}\left[\kappa\varepsilon n^2+n^2\ln\left(\frac{n}{\sqrt{e}}\right)\right]A_\mathrm{QD},
\end{split}
\end{eqnarray}
where $r_\mathrm{in}$ and $r_\mathrm{out}$ denote the inner and outer radii corresponding to the vortex core and droplet boundary, respectively. The first term, proportional to $S^2$, accounts for the kinetic energy associated with the quantized phase winding around the vortex core, naturally vanishing for ground-state droplets $(S=0)$. The second term encompasses the contributions from the nonlocal QQIs and the LHY quantum correction. Here, $\varepsilon = \int \mathrm{d}\mathbf{r}\, R(\mathbf{r}) = -\pi/b^3 = -\pi$ represents the integrated kernel associated with the quadrupolar interaction potential, and \( A_\mathrm{QD} = N/n \) defines the droplet's area in terms of its total norm $N$ and peak density $n$. Thus, this unified expression facilitates the consistent analysis of both ground-state and vortex-state QDs.

Minimizing the total energy given by Eq. (\ref{Eq:E}) with respect to $n$ yields the equilibrium density $n_e$ and corresponding  equilibrium area $A_e$:

\begin{equation}\label{Eq:ne}
	n_e=\exp\left({-\kappa \epsilon-\frac{1}{2}}\right), \quad A_e=N\exp\left(\kappa \epsilon + \frac{1}{2}\right) .
\end{equation}

Similarly, under the constant-density approximation, the chemical potential $\mu$ can be expressed as

\begin{equation}
\begin{split}
	\mu=&\frac{\pi S^2 n}{N}\ln\left(\frac{r_{\mathrm{out}}}{r_{\mathrm{in}}}\right) + \kappa\varepsilon n + n\ln n \\
	& = \frac{\pi S^2 n}{N}\ln\sqrt{1+\frac{N}{n\pi r_1^2}} +\kappa\varepsilon n + n \ln n.
\end{split}
\end{equation}

In the large-$N$ limit, this simplifies to

\begin{eqnarray}\label{Eq:mu}
	\mu\approx\kappa\varepsilon n + n\ln n.
\end{eqnarray}
Substituting the equilibrium density $n_e$ into the above equation, we obtain the equilibrium chemical potential as

\begin{eqnarray}\label{Eq:mue}
	\mu_e=-\frac{1}{2\sqrt{e}}e^{-\kappa\varepsilon}.
\end{eqnarray}

Importantly, the above analytical results show no explicit dependence on vorticity-related parameters. Thus, within the TF approximation, the equilibrium density $n_e$ and chemical potential $\mu_e$ remain invariant for both ground-state and vortex-state droplets. Conversely, the droplet area scales linearly with the total norm $N$, $A_e=N/n_e$, reflecting the droplet's incompressibility.

The TF approximation inherently neglects kinetic energy contributions, leading to deviations from numerical results, particularly at lower particle numbers. Nevertheless, numerical simulations exhibit increasing agreement with TF predictions as $N$ grows, validating the analytical expressions in the asymptotic regime. Detailed comparisons between theoretical predictions and numerical solutions are provided in Figs. \ref{Properties_GS} and \ref{Properties_VS}.

\section{Stationary properties and stability of quantum droplets}

Stationary QD solutions to Eq.~(\ref{Eq:GPE3}) are obtained numerically by means of the imaginary-time evolution method (ITM), which reliably converges to energy-minimizing states. The dynamical stability of these solutions is subsequently verified by subjecting them to direct simulations with random perturbations of relative amplitude at the 1\% level.

As the initial condition for the iterative computation, we use the following ansatz:

\begin{eqnarray}\label{Eq.11}
	\phi^{(0)}(x,y)=A\tilde{r}^S\exp\left(-\tilde{r}^2/r_0^2+iS\tilde{\theta}\right),
\end{eqnarray}
where $A=1$ and $r_0=10$ specify the amplitude and width of the trial function. The coordinates $(\tilde{r}, \tilde{\theta})=\big[\sqrt{x^2 + \beta^2 y^2}, \arctan(\beta y/x)\big]$ incorporate an anisotropy parameter $\beta=1.5$, which allows for the generation of elliptically deformed states. This ansatz facilitates the construction of both isotropic and anisotropic QDs, with or without embedded vorticity.

Once stationary profiles are computed, they are characterized by three key physical quantities: (i) the peak density, defined as $I_{\mathrm{max}}=\max|\phi(\mathbf{r})|^2$; (ii) the chemical potential $\mu$, extracted from the stationary equation (\ref{Eq:Stationary}); and (iii) the effective area $A_{\mathrm{eff}}$, which quantifies the spatial extent of the quantum droplet and is defined as

\begin{eqnarray}\label{Eq:Aeff}
	A_{\mathrm{eff}}=\frac{\left(\int|\psi|^2\mathrm{d}\textbf{r}\right)^2}{\int|\psi|^4\mathrm{d}\textbf{r}}. 
\end{eqnarray}

\subsection{Ground-state droplets: Density saturation and area-particle scaling}

\begin{figure*}[htbp]
	\centering
	\includegraphics[scale=0.45]{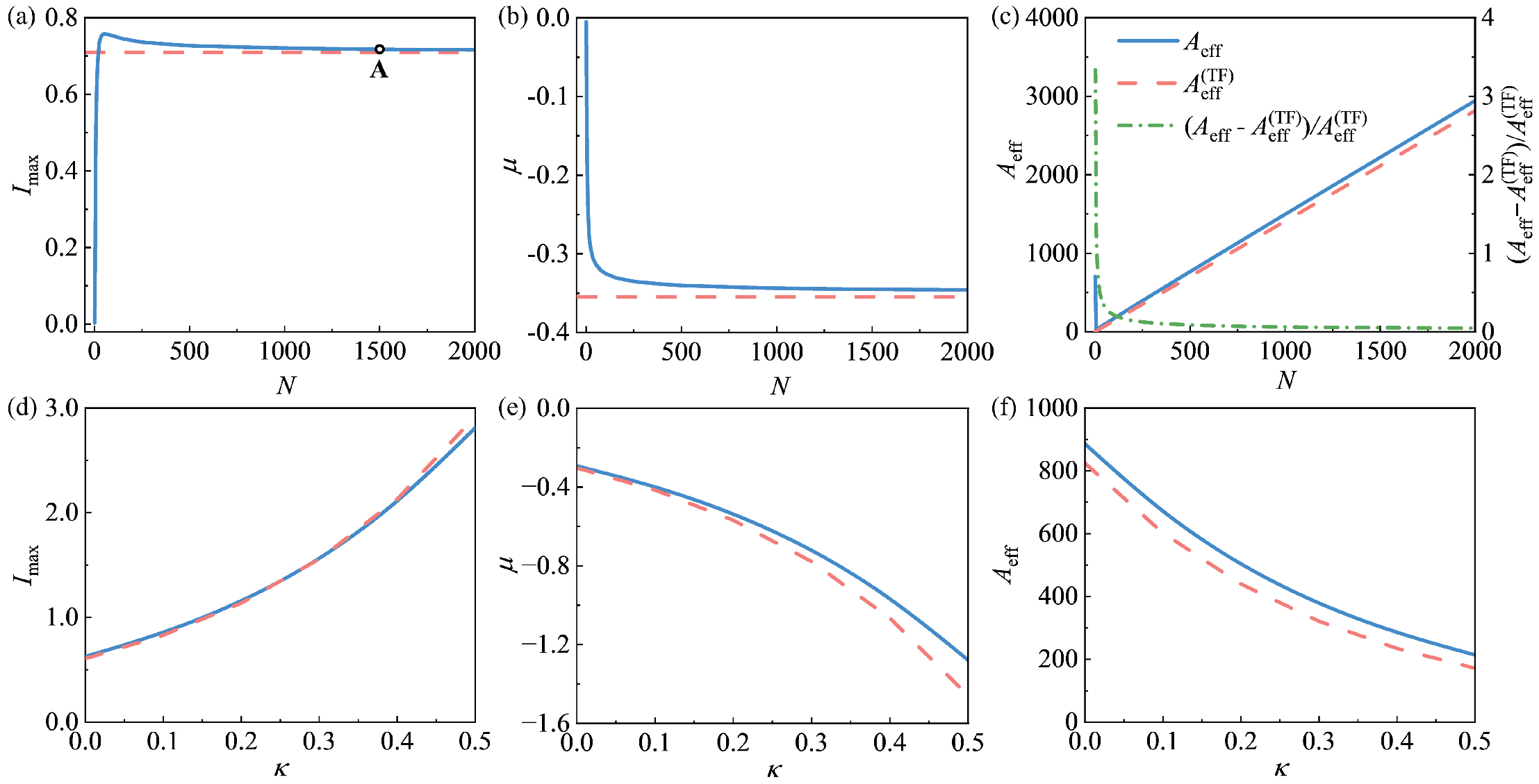}
	\caption{(Color online) Stationary properties of ground-state quantum droplets (QDs) under attractive QQIs and their dependence on the quadrupolar interaction strength. (a–c) Peak density $I_{\mathrm{max}}$, chemical potential $\mu$, and effective area $A_{\mathrm{eff}}$ as functions of the total norm $N$ at fixed quadrupolar strength $\kappa = 0.05$. With increasing $N$, $I_{\mathrm{max}}$ and $\mu$ gradually saturate, while $A_{\mathrm{eff}}$ grows approximately linearly in the large-$N$ regime, consistent with the incompressibility characteristic of quantum droplets. (d-f) Variation of $I_{\mathrm{max}}$, $\mu$, and $A_{\mathrm{eff}}$ as functions of $\kappa$ at fixed norm $N = 500$. As $\kappa$ increases, $I_{\mathrm{max}}$ increases and $A_{\mathrm{eff}}$ decreases, indicating stronger spatial localization of the droplets. Meanwhile, the chemical potential $\mu$ exhibits a monotonically decreasing trend. The red dashed curves indicate the analytical predictions based on the Thomas--Fermi approximation derived in Eqs. (\ref{Eq:ne}) and (\ref{Eq:mue}), and valid primarily in the large-$N$ regime.} \label{Properties_GS}
\end{figure*}

Figures \ref{Properties_GS}(a)–\ref{Properties_GS}(c) show the dependence of $I_{\mathrm{max}}$, $\mu$, and $A_{\mathrm{eff}}$ on the total norm $N$ for attractive QQIs with $\kappa = 0.05$. As shown in Fig.~\ref{Properties_GS}(a), the peak density increases with $N$ at first, reflecting enhanced self-localization, and then saturates near $I_{\mathrm{max}} \approx 0.715$, signaling the formation of a flat-topped droplet profile. This saturation behavior indicates the balance between attractive QQIs and repulsive LHY quantum pressure. Moreover, in the large-$N$ regime, the numerical results asymptotically approach the TF prediction (red dashed line), confirming the validity of the TF approximation for sufficiently high particle numbers.

In Fig.~\ref{Properties_GS}(b), the chemical potential $\mu$ decreases monotonically with $N$, satisfying the Vakhitov–Kolokolov (VK) stability criterion $d\mu/dN < 0$. At large $N$, $\mu$ approaches the theoretical value predicted by Eq. (\ref{Eq:mue}), as indicated by the dashed line.

The effective area $A_{\mathrm{eff}}$ in Fig.~\ref{Properties_GS}(c) exhibits a nonmonotonic behavior as a function of the total norm $N$: it decreases rapidly in the small-$N$ regime and then increases approximately linearly for larger $N$. This trend reflects a characteristic \textit{gas-to-liquid crossover} \cite{Chomaz2016,Cabrera2018}. When $N$ is small, the quantum pressure dominates, leading to an extended low-density state with gas-like features. As $N$ increases, the quadrupolar attraction gradually overcomes the delocalizing effect of kinetic energy, resulting in a self-bound droplet with a flat-topped density profile. This crossover marks the transition to an incompressible quantum liquid state, as evidenced by the subsequent linear scaling $A_{\mathrm{eff}} \propto N$, consistent with the TF prediction $A_{\mathrm{e}} = N/n_{\mathrm{e}}$ (red dashed line).

Although the absolute deviation between the numerical results and the TF approximation appears to increase with $N$, the relative error (see green dash-dotted curve) actually decreases, indicating that the numerical results approach the TF prediction in the large-$N$ limit.

While ground-state droplets are generally considered to exist for arbitrarily small $N$, our numerical results suggest that, below a certain threshold, stable localized solutions become difficult to realize, either due to insufficient binding energy or significant quantum fluctuations. Therefore, in practice, a lower bound for droplet stability may still emerge, albeit without a sharp critical point as in the vortex case (see Sec.~\ref{sub-sec_VS}).

\subsection{Tuning droplet properties via quadrupolar interactions}

Figures \ref{Properties_GS}(d)–\ref{Properties_GS}(f) illustrate how the properties of ground-state QDs vary with quadrupolar interaction strength $\kappa$ at a fixed total norm $N = 500$. As shown in Fig.~\ref{Properties_GS}(d), the peak density $I_{\max}$ increases monotonically with the quadrupolar interaction strength $\kappa$. This behavior indicates that stronger quadrupole-quadrupole interactions lead to tighter spatial confinement of the droplet, resulting in a higher central density. The excellent agreement between the numerical results (blue solid curve) and the TF prediction (red dashed curve) suggests that, in this regime, the system is well described by the balance between nonlocal attractive interactions and repulsive quantum fluctuations. The steady increase of $I_{\max}$ confirms that quadrupolar attraction dominates the droplet formation as $\kappa$ grows. 

Figure \ref{Properties_GS}(e) shows that the chemical potential $\mu$ decreases monotonically with increasing $\kappa$, becoming more negative. This trend reflects stronger binding of the droplet as the quadrupolar attraction intensifies. The deepening of the chemical potential implies that the system enters a more self-bound state, with enhanced internal cohesion. Although the numerical and analytical results follow a similar trend, a slight deviation is observed at larger $\kappa$, indicating the increasing role of kinetic and higher-order corrections beyond the TF approximation.

In Fig.~\ref{Properties_GS}(f), the effective area $A_{\mathrm{eff}}$ decreases monotonically with increasing $\kappa$. This behavior demonstrates that the droplet becomes more spatially localized as the quadrupolar interactions strengthen. A larger $\kappa$ compresses the droplet due to the enhanced nonlocal attraction, overcoming the spreading effect of quantum fluctuations. The comparison with the TF prediction again shows good agreement, confirming that the size of the droplet can be effectively tuned by controlling $\kappa$.

\subsection{Vortex-state droplets: Vorticity-induced modifications}\label{sub-sec_VS}

Previous studies have systematically explored the existence, structural characteristics, and dynamical stability of vortex quantum droplets (vQDs) in two-dimensional systems, uncovering fundamental insights into the influence of vorticity on droplet morphology and robustness~\cite{Li2018}. Anisotropic vortex quantum droplets have previously been reported in dipolar BECs, where dipole–dipole interactions give rise to elliptical density profiles and nontrivial angular momentum distributions~\cite{Li2024}. Theoretical investigations have further revealed distinctive rotational dynamics and stability conditions inherent to self-bound vortex states, emphasizing the critical role of topological constraints~\cite{Cidrim2018,Dong2021}. More recently, vQDs stabilized by competing nonlinear interactions have been systematically examined, highlighting the impact of nonlinear competition on their structural profiles and stability thresholds~\cite{Chen2024}. Building on this foundation, we extend the analysis to vQDs with topological charge $S = 1$.

\begin{figure*}[htbp]
	\centering
	\includegraphics[scale=0.45]{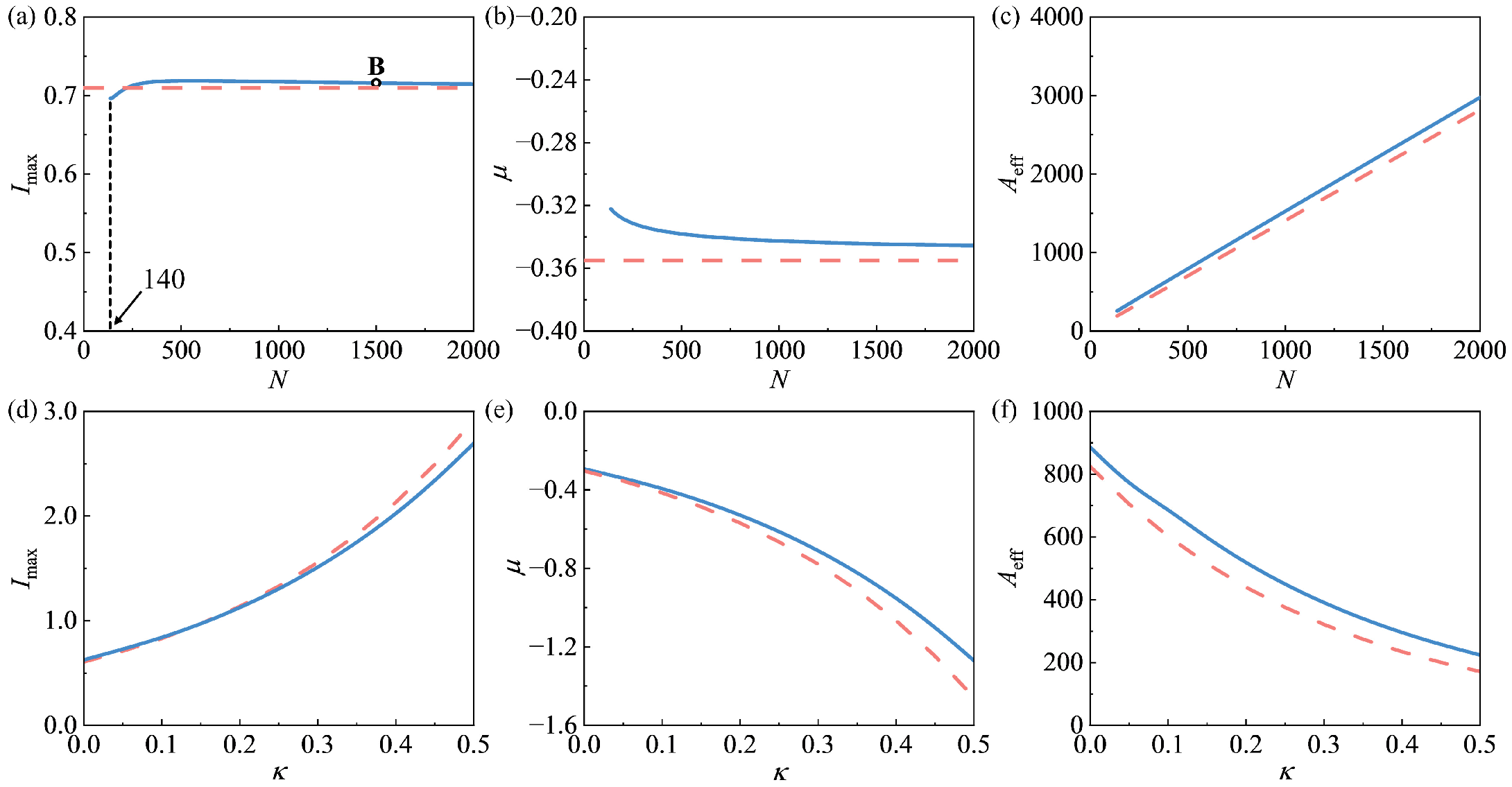}
	\caption{(Color online) Stationary properties of vortex quantum droplets (vQDs) under attractive quadrupole–quadrupole interactions (QQIs) and the influence of the interaction strength. (a1–a3) Peak density $I_{\mathrm{max}}$, chemical potential $\mu$, and effective area $A_{\mathrm{eff}}$ as functions of total norm $N$ at fixed quadrupolar strength $\kappa = 0.05$. The peak density and chemical potential exhibit gradual saturation as $N$ increases. Compared to the ground-state droplets, although it’s not obvious, vQDs feature a slightly larger effective area $A_{\mathrm{eff}}$ at the same $N$, resulting from the central density depletion that displaces particles outward. (b1–b3) Dependence of $I_{\mathrm{max}}$, $\mu$, and $A_{\mathrm{eff}}$ on the quadrupolar interaction strength $\kappa$ at a fixed norm $N = 500$. With increasing $\kappa$, $I_{\mathrm{max}}$ increases while $\mu$ decreases monotonically, indicating enhanced self-binding. Meanwhile, $A_{\mathrm{eff}}$ shrinks due to the stronger attractive interactions. Red dashed lines represent theoretical predictions from the TF approximation, using Eqs. (\ref{Eq:ne}) and (\ref{Eq:mue}).} \label{Properties_VS}
\end{figure*}

Figure~\ref{Properties_VS} presents a comprehensive analysis of the key physical properties of vQDs, focusing on their dependence on the total particle number $N$ (top row) and the quadrupolar interaction strength $\kappa$ (bottom row). In comparison to their ground-state counterparts shown in Fig.~\ref{Properties_GS}, the introduction of vorticity leads to subtle yet significant modifications in the peak density, chemical potential, and effective area. A notable distinction between vQDs and fundamental droplets lies in the existence threshold: while ground-state droplets persist even at vanishingly small $N$, stable vortex states only emerge above a critical norm, approximately $N_{\mathrm{cr}} \approx 140$. This threshold behavior reflects the additional energetic cost required to sustain a quantized vortex core.

In Fig.~\ref{Properties_VS}(a), the peak density $I_{\max}$ of the vQD increases rapidly for small $N$ and saturates around 0.71 at large $N$, matching the saturation value of the fundamental droplets [cf.~Fig.~\ref{Properties_GS}(a)]. This indicates that the presence of a vortex does not significantly alter the peak density in the high-$N$ regime. Figure~\ref{Properties_VS}(b) shows a monotonic decrease in the chemical potential $\mu$ with increasing $N$, asymptotically approaching –0.35. This behavior, similar to that in Fig.~\ref{Properties_GS}(b), signals enhanced self-binding of the droplet with increasing norm. Notably, the TF prediction closely follows the numerical data for large $N$, reaffirming its validity in the high-density regime. Figure~\ref{Properties_VS}(c) demonstrates that the effective area $A_{\mathrm{eff}}$ increases linearly with $N$, consistent with the incompressibility characteristic of quantum droplets. Compared to the fundamental case [Fig.~\ref{Properties_GS}(c)], the effective area of vQDs is slightly larger at fixed $N$ due to the central density depletion induced by the vortex. This is also reflected in the slightly greater slope of the linear scaling. The TF approximation captures the overall trend well, with minor deviations in the low-$N$ regime attributed to kinetic effects not accounted for in the TF model.

The bottom row of Fig.~\ref{Properties_VS} explores the influence of quadrupolar interaction strength $\kappa$ at fixed norm $N=500$. As shown in Fig.~\ref{Properties_VS}(d), $I_{\max}$ increases monotonically with $\kappa$, but remains consistently lower than that of the corresponding ground-state droplets [Fig.~\ref{Properties_GS}(d)], particularly at large $\kappa$. This difference arises from the kinetic pressure associated with the vortex core, which counteracts the compressive effect of quadrupolar attraction. Figure\ref{Properties_VS}(e) shows that the chemical potential decreases monotonically with $\kappa$, closely mirroring the trend observed in the fundamental case [Fig.~\ref{Properties_GS}(e)]. This suggests that the vortex kinetic energy remains small compared to the contributions from QQI and quantum fluctuations. Finally, Fig.~\ref{Properties_VS}(f) reveals a monotonic decrease in $A_{\mathrm{eff}}$ with increasing $\kappa$, indicating stronger spatial confinement of the droplet. However, the degree of compression is slightly reduced compared to Fig.~\ref{Properties_GS}(f), again due to the incompressible core introduced by vorticity. The TF prediction slightly underestimates the effective area of vQDs, particularly at low $\kappa$, owing to its neglect of vortex-related kinetic energy.

These results confirm that vortex quantum droplets inherit the incompressibility and stability of fundamental droplets, while exhibiting distinct quantitative corrections due to their nontrivial topological structure. The TF approximation remains valid in the large-$N$ or small-$\kappa$ regimes, but deviations arise at higher interaction strengths where vortex-induced kinetic energy becomes significant.

\begin{figure}[htbp]
	\centering
	\includegraphics[scale=0.38]{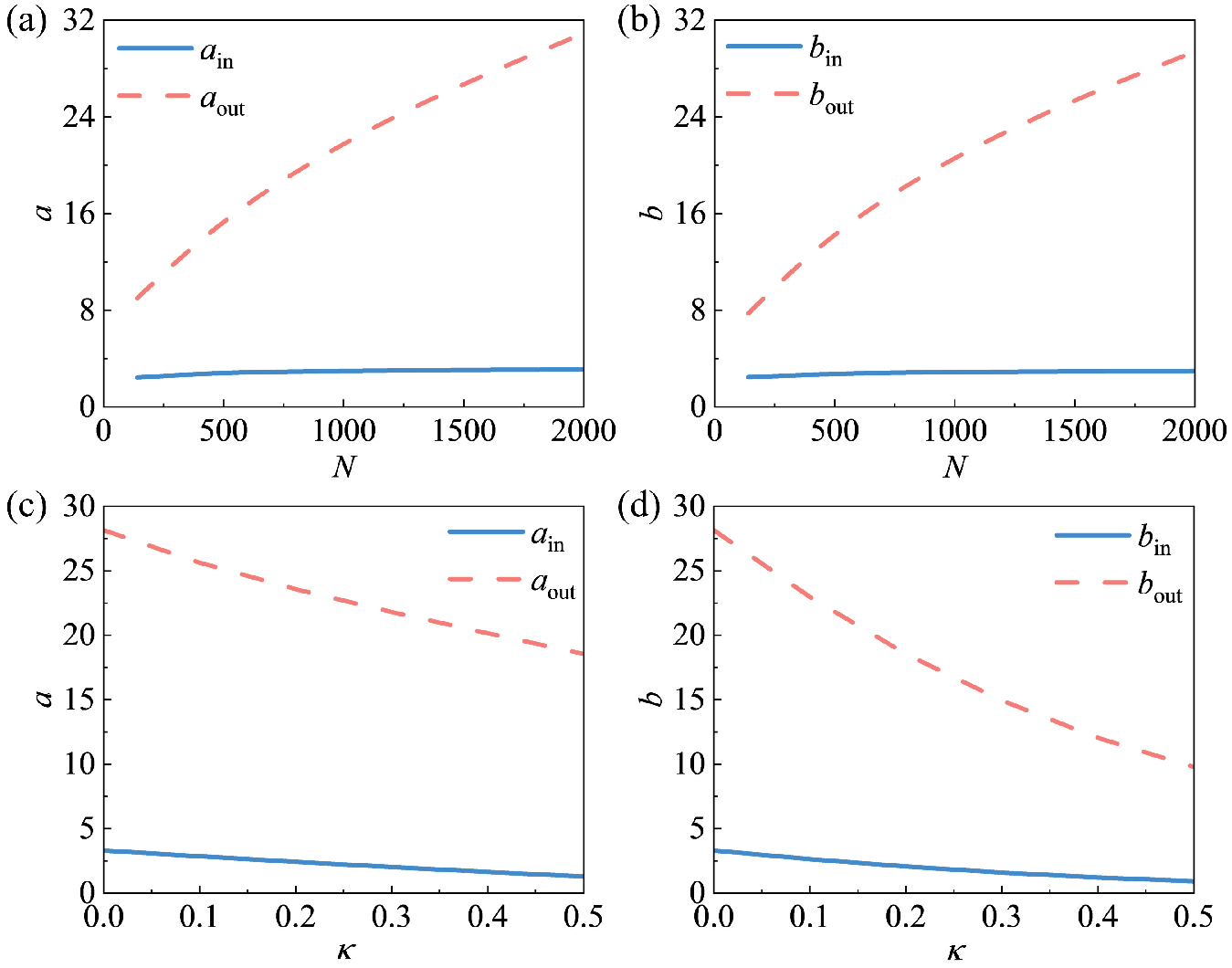}
	\caption{(Color online) Semi-major ($a$) and semi-minor ($b$) axes of the inner vortex core and outer boundary of vortex quantum droplets. Panels (a) and (b) show the dependence on total norm $N$ at fixed quadrupolar strength $\kappa = 0.05$, while panels (c) and (d) illustrate the dependence on $\kappa$ at fixed $N = 1500$. As $N$ increases, the outer boundary expands monotonically, whereas the core size remains nearly unchanged. In contrast, increasing $\kappa$ leads to overall droplet contraction, with the outer boundary shrinking more noticeably than the vortex core. The anisotropic nature of QQIs is evident in (c,d), where the semi-minor axis ($b$) exhibits a more pronounced reduction than the semi-major axis ($a$). In all panels, blue solid lines denote the inner boundary dimensions, and red dashed lines denote the outer boundary dimensions.} \label{Morphology_VS}
\end{figure}

Further elucidating the anisotropic character of vortex quantum droplets, we recall that the anisotropic nature of the QQIs induces elliptical morphologies in both ground-state and vortex-state quantum droplets. Similar anisotropic vortex deformations driven by dipolar interactions have been recently reported in dipolar BECs~\cite{Li2024a}, further validating the universality of anisotropy in multipolar quantum fluids. Figure \ref{Morphology_VS} quantifies this ellipticity by tracking the semi-major ($a$) and semi-minor ($b$) axes of both the inner vortex core and outer droplet boundary under variations of the total norm $N$ and quadrupolar interaction strength $\kappa$.

Panels (a) and (b) in Fig.~\ref{Morphology_VS} reveal that as $N$ increases, the vortex core dimensions (inner axes, $a_{\mathrm{in}}$ and $b_{\mathrm{in}}$) remain nearly constant, while the outer boundary axes ($a_{\mathrm{out}}$ and $b_{\mathrm{out}}$) exhibit monotonic growth. This stark contrast underscores the incompressible nature of the quantum liquid phase: increased particle number expands the droplet’s spatial footprint without compressing the central topological defect, consistent with the linear scaling of effective area $A_{\mathrm{eff}}\propto N$ observed in Fig.~\ref{Properties_VS}(c). 

Conversely, Panels (c) and (d) in Fig.~\ref{Morphology_VS} demonstrate that strengthening quadrupolar interactions (increasing $\kappa$) compresses the entire droplet anisotropically. Crucially, the inner and outer boundaries respond dissimilarly: while the vortex core contracts only mildly, the outer boundary undergoes significant shrinkage, with the semi-minor axis $b_{\mathrm{out}}$ exhibiting greater sensitivity to $\kappa$ than the semi-major axis $a_{\mathrm{out}}$. This differential compression---where $|\Delta b_{\mathrm{out}}|>|\Delta a_{\mathrm{out}}|\gg|\Delta b_{\mathrm{in}}|\approx|\Delta a_{\mathrm{in}}|$---directly manifests the anisotropy inherent to quadrupole-quadrupole interactions (QQIs). The enhanced contraction along the minor axis reflects the preferential alignment of attractive QQI forces perpendicular to the quadrupole’s symmetry axis (here, $x$), thereby accentuating the droplet’s elliptical deformation. These observations provide unambiguous spatial evidence that QQIs not only tune the droplet’s size [as captured by $A_{\mathrm{eff}}$ in Fig.~\ref{Properties_VS}(f)] but also actively reshape its aspect ratio, highlighting a key distinction from isotropic quantum droplets.

While our analysis focuses on $S=1$, the TF predictions for bulk observables (flat-top density, $\mu(N)$ saturation, and $A_{\mathrm{eff}}\!\propto\!N$) are expected to remain valid for $S>1$ at large $N$, since they are governed by the bulk balance between quadrupolar attraction and LHY repulsion. The centrifugal kinetic energy near the core scales as $\Delta E_{\mathrm{kin}}\!\propto\!(S^2 N/R^2)\ln(R/\xi)$, where $\xi$ is the \emph{healing length} (introduced here as the characteristic vortex--core size). This scaling raises the minimal norm for stability (up to logarithmic corrections) and narrows the stability window with increasing $S$. As a result, higher-order vQDs are more prone to azimuthal instabilities or splitting in anisotropic QQI settings. Related isotropic-MQQI results have reported finite stability for $S=2$ and constructed $S\!\ge\!3$ states at larger norms,
consistent with this scaling-based expectation~\cite{Yang2024}.

Figure \ref{Examples_GS_VS} demonstrates typical examples of stable ground-state and vortex-state QDs formed under QQIs, with fixed parameters $N = 1500$ and $\kappa = 0.05$, as indicated by points A and B in Figs.~\ref{Properties_GS} and \ref{Properties_VS}. The top panels (a,b) show the stationary density profiles, while the bottom panels (c,d) display their time evolution under weak random perturbations. The ground-state QD exhibits a nearly uniform, flat-topped density profile with sharp edges, characteristic of an incompressible quantum liquid. In contrast, the vQD features a pronounced density dip at the center, as expected from the presence of a quantized vortex. The inset of Fig.~\ref{Examples_GS_VS}(b) confirms the vortex phase structure, revealing a $2\pi$ phase winding around the central singularity.

\begin{figure}[htbp]
	\centering
	\includegraphics[scale=0.38]{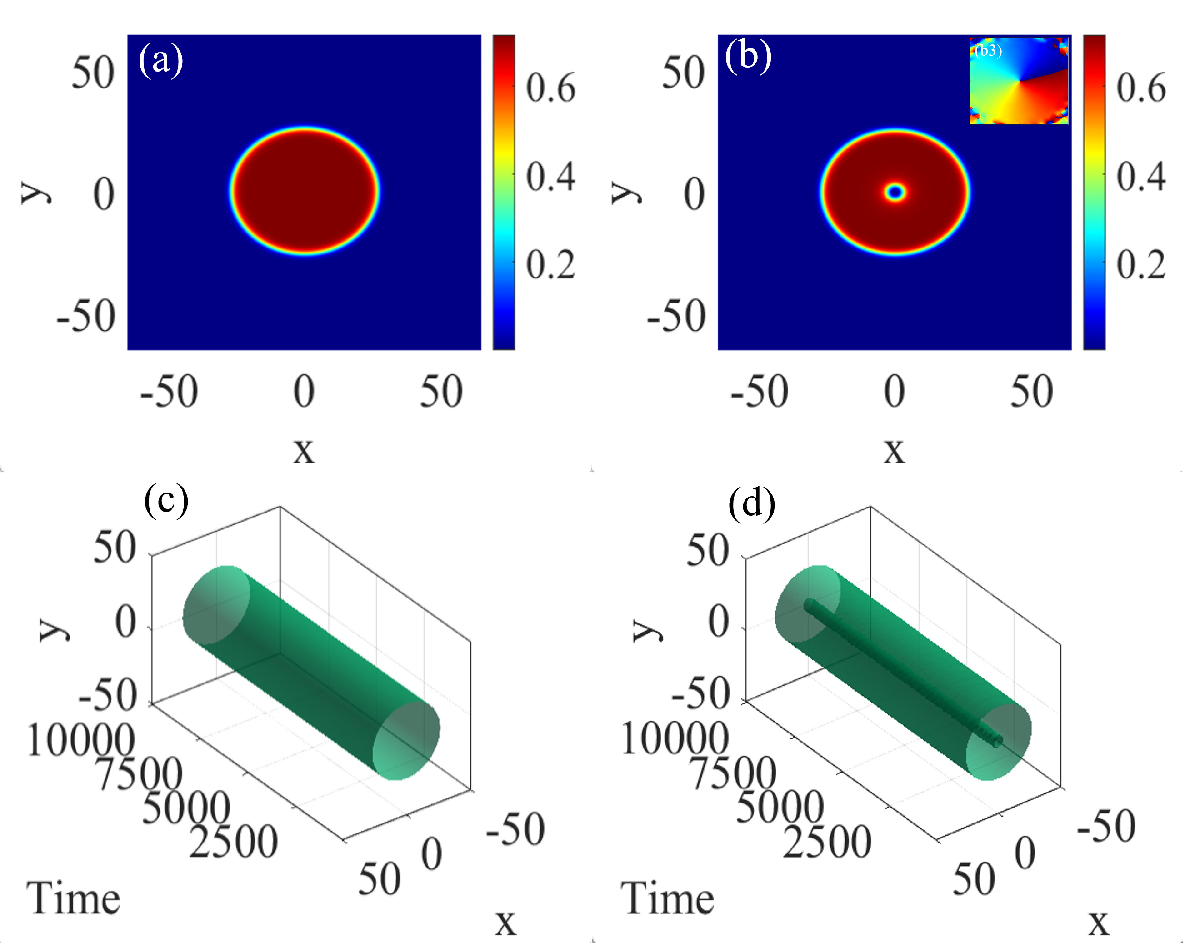}
	\caption{(Color online) Typical examples of dynamically stable ground-state and vortex-state quantum droplets (QDs) under attractive QQIs. (a1) Density distribution of the ground-state QD corresponding to point A in Fig.~\ref{Properties_GS}, with parameters $N=1500$ and $\kappa=0.05$. (b1) Density distribution of the vortex-state QD under the same parameters, corresponding to point B in Fig.~\ref{Properties_VS}. A clear central density depletion is observed due to the embedded vortex core. The inset in (b1) shows the corresponding phase pattern after real-time evolution, featuring a $2\pi$ phase winding that confirms the presence of a singly charged vortex. (a2) and (b2) show the results of real-time simulations with 1\% random perturbations added to the initial states shown in (a1) and (b1), respectively. The persistent density profiles over time confirm that both the ground-state and vortex-state QDs exhibit robust dynamical stability.} \label{Examples_GS&VS}
\end{figure}

Figures \ref{Examples_GS_VS}(c) and \ref{Examples_GS_VS}(d) present real-time simulations of perturbed ground-state and vortex-state QDs under weak (1\%) random noise. In both cases, the droplets remain structurally intact over the simulated time scale, indicating a degree of dynamical robustness against small perturbations. In particular, the vortex core in Fig.~\ref{Examples_GS_VS}(d) persists throughout the evolution, suggesting the possible stability of the topological structure. These numerical results support---but do not fully establish---the stability of the stationary states, and further investigations (e.g., Bogoliubov–de Gennes analysis or dynamical response under stronger perturbations) would be necessary to clarify the underlying stabilizing mechanisms.

\section{Collision dynamics of quantum droplets}

Collision dynamics provide crucial insights into the stability, interaction mechanisms, and topological characteristics of QDs. Recent experimental investigations have significantly advanced our understanding of these phenomena~\cite{Ferioli2019}, extending foundational insights into matter-wave interactions established in earlier studies~\cite{Nguyen2014}. Furthermore, theoretical analyses have explored collision dynamics and collective excitations in lower-dimensional quantum droplets, elucidating fundamental aspects such as dynamical stability, breathing modes, and intrinsic nonlinear wave phenomena~\cite{Astrakharchik2018}. To systematically examine head-on collisions, we prepare two droplets initially separated along the $x$-axis and impose opposite phase gradients ($\pm v$), thereby setting the droplets into motion. The initial wavefunction is given by:

\begin{equation}
	\psi(r,0)=\phi(r+x_0)e^{ivx}+\phi(r-x_0)e^{-ivx},
\end{equation}
where $x_0$ denotes the initial separation, and $v$ the incident velocity. The quadrupole moments are polarized perpendicular to the $x$-$y$ plane.

Throughout, the labels ``low/intermediate/high velocity" are phenomenological and chosen per sector: for ground-state QDs, $v=0.05$ already lies in the merger regime [Fig. \ref{Collision_GS}(a1–a4)], whereas for same-vorticity vortex QDs we use a smaller $v=0.015$ to cleanly reveal phase-induced repulsion as a bounce-off [Fig. \ref{Collision_VS}(c1–c3)]; at larger $v$ the kinetic influx overwhelms the core-mediated repulsion and leads to fragmentation or penetration.

\subsection{Ground-state collisions: Merging to quantum penetration}

\begin{figure*}[htbp]
	\centering
	\includegraphics[scale=0.5]{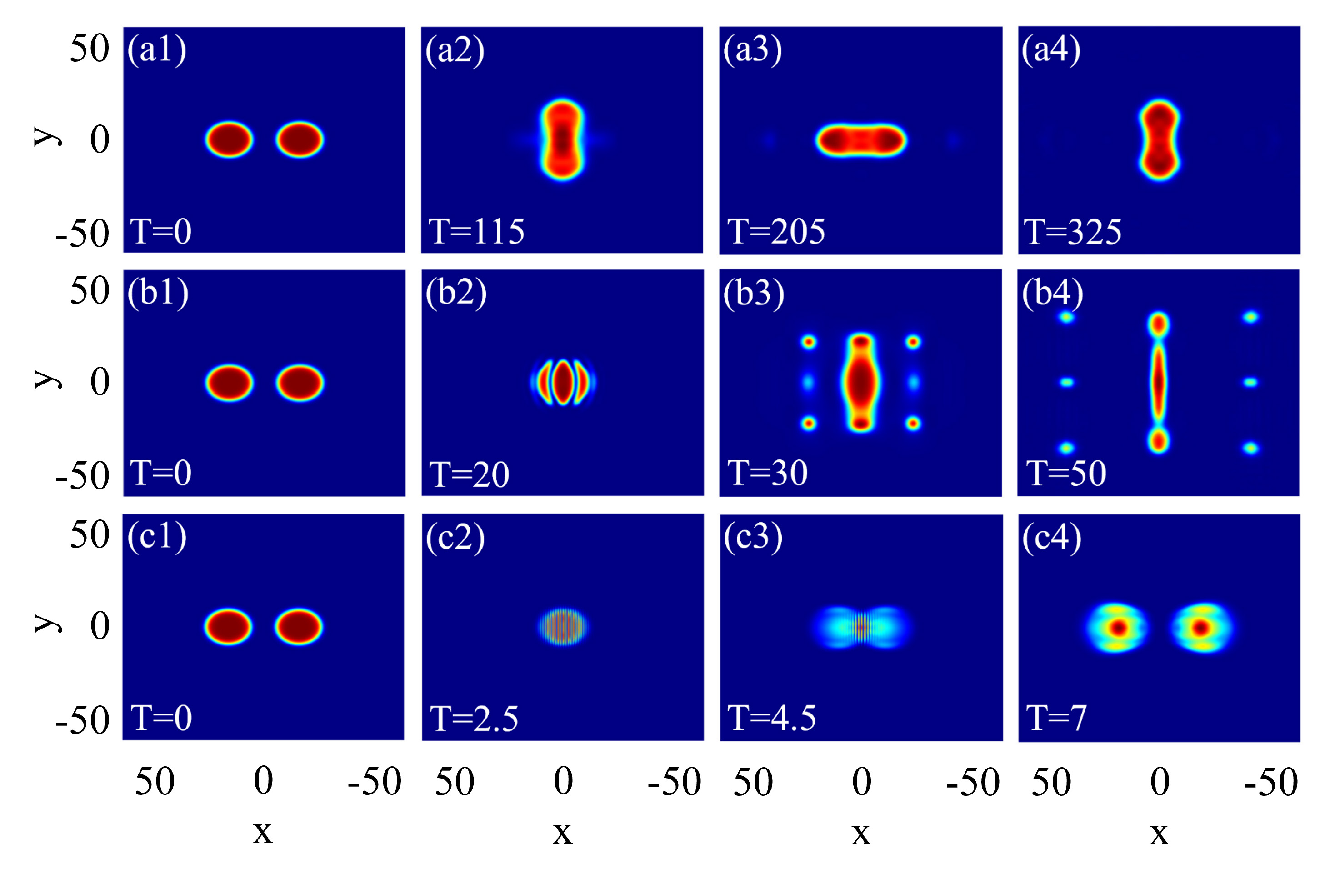}
	\caption{(Color online) Collision dynamics of ground-state quantum droplets (QDs) under attractive quadrupole–quadrupole interactions (QQIs) for different initial velocities. All simulations are performed with initial parameters $x_0=15$, $N=200$, and $\kappa=0.05$. (a1–a4) Low-velocity case ($v=0.05$): the droplets undergo an inelastic merger, coalescing into a single elongated droplet that exhibits sustained oscillations. (b1–b4) Intermediate velocity ($v=0.5$): the droplets approach, overlap briefly, and then scatter along the $y$-axis in a quasi-elastic manner, without significant loss of coherence. (c1–c4) High-velocity case ($v=3.0$): the droplets penetrate through each other with minimal distortion, displaying interference fringes characteristic of quantum coherence during the collision process.} \label{Collision_GS}
\end{figure*}

Figure~\ref{Collision_GS} presents the collision outcomes of ground-state QDs under varying initial velocities. At low velocity [$v = 0.05$, Figs.~\ref{Collision_GS}(a1–a4)], the droplets undergo inelastic merging, forming a single, oscillating entity. The resulting state exhibits radial breathing modes, indicating substantial kinetic energy dissipation and dominance of attractive interactions.

At intermediate velocity [$v=0.5$, Figs.~\ref{Collision_GS}(b1–b4)], the droplets briefly coalesce and then scatter perpendicularly along the $y$-axis, resembling quasi-elastic deflection. This $90^\circ$ scattering reflects anisotropic momentum redistribution and is characteristic of long-range quadrupolar interactions \cite{Hu2022}.

At high velocity [$v=3.0$, Figs.~\ref{Collision_GS}(c1–c4)], the droplets undergo quantum penetration, passing through each other with minimal distortion. Transient interference fringes appear during overlap, indicative of coherent phase interference. The overall integrity of each droplet is preserved, confirming that kinetic energy dominates over inter-droplet attraction in this regime.

\subsection{Vortex collisions: Vorticity preservation and fragmentation}

\begin{figure*}[htbp]
	\centering
	\includegraphics[scale=0.5]{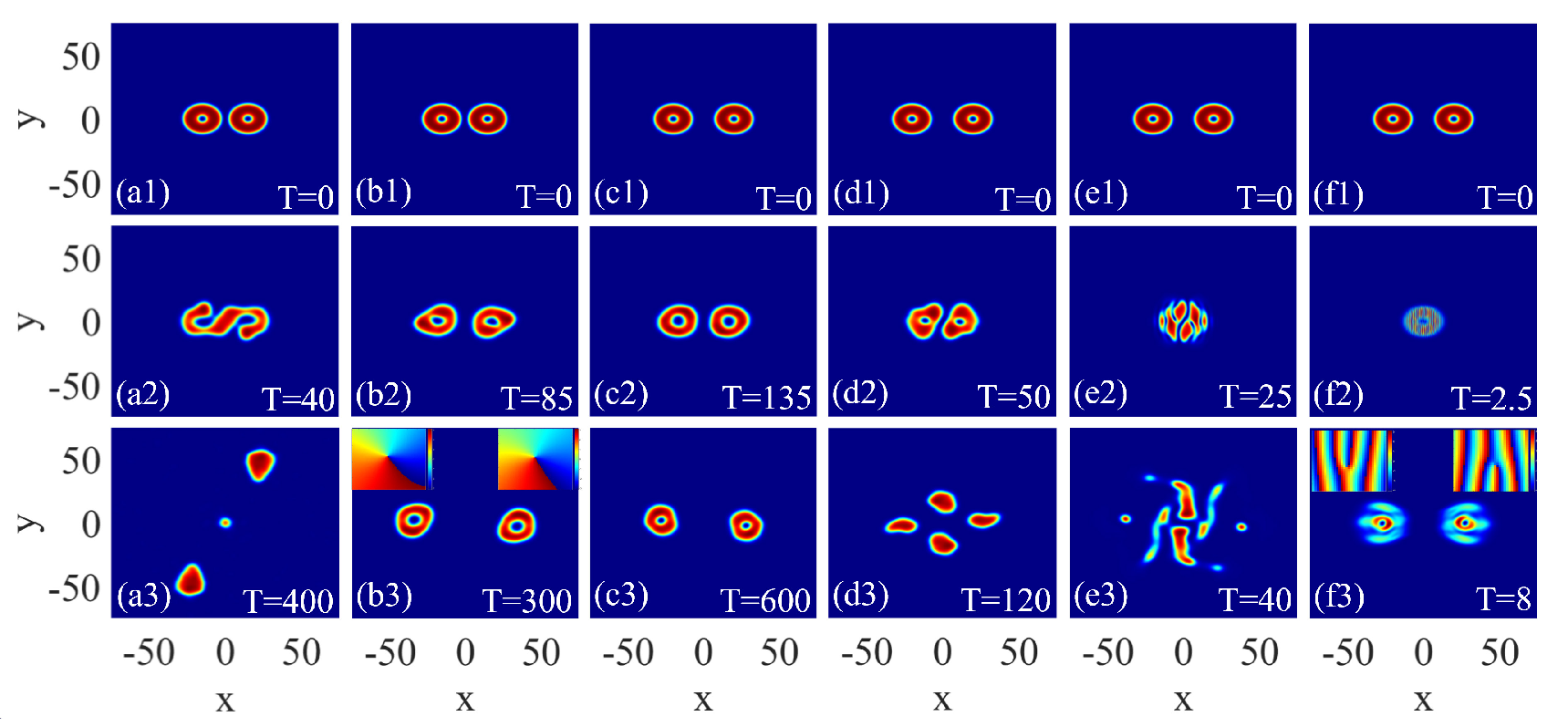}
	\caption{(Color online) Collision dynamics of vortex quantum droplets (vQDs) under attractive quadrupole–quadrupole interactions (QQIs), with parameters $N=300$ and $\kappa=0.05$. (a1–a3) Head-on collision between two vQDs with opposite topological charges ($S_1 = +1$, $S_2 = -1$) at zero velocity ($v = 0$). The droplets merge and subsequently fragment due to phase singularity annihilation. (b1–b3) Zero-velocity collision of two same-charge vQDs ($S_1 = S_2 = +1$), which exhibit mutual repulsion and maintain their internal phase structure. (c1–c3) At low velocity ($v = 0.015$), same-charge vQDs undergo quasi-elastic bouncing without significant deformation. (d1–d3) At moderate velocity ($v = 0.1$), the collision results in partial fragmentation and the generation of non-vortex daughter droplets, indicating loss of topological coherence. (e1–e3) For $v = 0.5$, complex interference patterns emerge, with partial retention of vorticity in the post-collision states. (f1–f3) At high velocity ($v = 3.0$), the vQDs fully penetrate each other while preserving their vortex structures, illustrating robust topological resilience under strong impact.} \label{Collision_VS}
\end{figure*}

Figure~\ref{Collision_VS} shows the dynamics of vortex QD collisions under varying velocities and topological configurations. At zero velocity, opposite-vorticity vQDs ($S_1 = 1$, $S_2 = -1$) merge and then fragment into multiple pieces [Figs.~\ref{Collision_VS}(a1–a3)], indicating instability due to destructive interference between opposing phase windings. In contrast, same-vorticity vQDs ($S_1 = S_2 = 1$) repel each other without coalescing [Figs.~\ref{Collision_VS}(b1–b3)], due to the effective repulsive interaction arising from their identical circulation and associated phase gradients. The phase diagram in Fig.~\ref{Collision_VS}(b3) confirms preservation of the vortex cores.

At low velocity [$v = 0.015$, Figs.~\ref{Collision_VS}(c1–c3)], same-vorticity vQDs approach and bounce off each other, demonstrating persistent phase-induced repulsion. At moderate velocity [$v = 0.1$, Figs.~\ref{Collision_VS}(d1–d3)], collision triggers vortex breakdown and results in fragmentation into four smaller non-vortex droplets, indicating destruction of topological coherence.

At higher velocities [$v = 0.5$ and $v = 3.0$, Figs.~\ref{Collision_VS}(e1–f3)], prominent interference fringes appear at the moment of overlap. At $v = 0.5$, the droplets exhibit partial vortex preservation and form transient filamentary structures. At $v = 3.0$, the droplets clearly penetrate one another, and phase diagram [Fig.~\ref{Collision_VS}(f3)] confirms the survival of the vortex structure.

These results demonstrate that QD collisions are governed by a competition between kinetic energy, long-range quadrupolar attraction, and phase coherence. Ground-state QDs transition from inelastic merging to elastic scattering and eventually to penetration with increasing velocity. Vortex-state QDs exhibit richer behaviors, including phase-induced repulsion, vortex breakdown, and topologically preserved tunneling. The ability to control collision outcomes via velocity and vorticity highlights the potential of quadrupolar droplets for studying nonlinear dynamics and topological quantum fluids.

\section{Conclusion}

In this work, we have systematically investigated the existence, structural features, and collision dynamics of quantum droplets (QDs) in a two-component quadrupolar Bose--Einstein condensate confined to a quasi-two-dimensional geometry. By incorporating nonlocal quadrupole--quadrupole interactions (QQIs) and beyond-mean-field Lee--Huang--Yang (LHY) corrections into an extended Gross--Pitaevskii framework, we have explored both ground-state and vortex-state droplets.

From a theoretical perspective, we applied a Thomas--Fermi (TF) approximation to obtain analytical predictions for the equilibrium density, chemical potential, and effective area of flat-top droplets. These expressions reveal that the droplet density is independent of total particle number, while the area scales linearly with it. The analytical results show good agreement with the numerical findings in the large-$N$ regime.

On the numerical side, we have solved the extended Gross–Pitaevskii equation to obtain stationary and dynamically stable droplet solutions. For both fundamental and vortex QDs, we observed saturation of peak density and chemical potential with increasing particle number, as well as linear scaling of effective area, confirming the incompressible nature of these self-bound states. Notably, we have found that vortex quantum droplets exhibit pronounced elliptical deformation due to anisotropic QQIs, with the droplet’s aspect ratio sensitively dependent on the total particle number and quadrupolar interaction strength. 

Real-time simulations under weak perturbations confirm their dynamical robustness, emphasizing the stabilizing role of QQIs and quantum pressure arising from both kinetic energy and beyond-mean-field LHY corrections.

The analysis of collision dynamics reveals distinct interaction scenarios governed by the interplay between velocity, interaction strength, and topological charge. Ground-state QDs display transitions from inelastic merging to quasi-elastic scattering and quantum penetration with increasing impact velocity. Vortex-state QDs demonstrate richer dynamics, including phase-induced repulsion, fragmentation, and the preservation of vortex cores at high velocities.

Overall, our findings deepen the understanding of quadrupolar quantum droplets and establish a comprehensive theoretical and computational framework for analyzing their structural and dynamical properties. The demonstrated tunability of vortex droplet morphologies through particle number and quadrupolar interaction strength provides new insights into the manipulation of anisotropic quantum fluids. These results open new avenues for the exploration of anisotropic quantum liquids, topological excitations, and their controlled manipulation in ultracold atomic systems. Future directions include studying multi-vortex interactions, engineering droplet dynamics using external potentials, and experimentally realizing quadrupolar droplets using polar molecules.

\subsection*{CRediT authorship contribution statement}

Wei-qi Xia: Software, Investigation, Data curation, Writing – original draft. Xiao-ting Zheng: Software, Investigation, Data curation, Writing – original draft. Xiao-wei Chen: Investigation, Validation. Gui-hua Chen: Conceptualization, Writing – review \& editing, Supervision.

\subsection*{Data Availability Statement}

The data supporting this study's findings are available within the article.

\subsection*{Declaration of competing interest}

The authors declare that they have no known competing financial interests or personal relationships that could have appeared to influence the work reported in this paper.

\begin{acknowledgments}
 G.-h.~Chen appreciates useful discussions with Dr. Yongyao Li (Foshan University). The work is supported by the Guangdong Basic and Applied Basic Research Foundation (Grant No.~2024A1515010710). We are grateful for the financial support provided by the organization, which has been instrumental in carrying out this research.
\end{acknowledgments}

\end{document}